\documentclass[twocolumn,showpacs,preprintnumbers,amsmath,amssymb,pre]{revtex4}

\usepackage{graphicx}
\usepackage{dcolumn}
\usepackage{bm}
\usepackage{enumerate}
\usepackage{amsmath}
\usepackage{color}

\begin{document}

\title{Subdiffusion with particle immobilization process described by differential equation with Riemann--Liouville type fractional time derivative}

\author{Tadeusz Koszto{\l}owicz}
 \email{tadeusz.kosztolowicz@ujk.edu.pl}
 \affiliation{Institute of Physics, Jan Kochanowski University,\\
         Uniwersytecka 7, 25-406 Kielce, Poland}

\date{\today}

\begin{abstract}
An equation describing subdiffusion with possible immobilization of particles is derived by means of the continuous time random walk model. The equation contains a fractional time derivative of Riemann--Liouville type which is a differential-integral operator with the kernel defined by the Laplace transform. We propose the method for calculating the inverse Laplace transform providing the kernel in the time domain. In the long time limit the subdiffusion--immobilization process reaches a stationary state in which the probability density of a particle distribution is an exponential function. 
\end{abstract}

\maketitle

\section{Introduction}

In diffusion process particles can be eliminated from further diffusion in different ways. There may be a particle decay due to a reaction when it meets other molecules. Since the particle disappears, the probability density $P(x,t)$ that the particle is at a point $x$ in time $t$ is not normalized,
\begin{equation}\label{eq1}
\int_{-\infty}^\infty P(x,t)dx<1.
\end{equation}
Another process that eliminates a particle from further diffusion is the permanent immobilization of the particle. Both processes mentioned above can occur in the diffusion of antibiotic molecules in a bacterial biofilm. One of defense mechanisms is to disintegrate the antibiotic molecules, the process can be described by diffusion-reaction equations. In the other one bacteria can thicken the biofilm immobilizing antibiotic molecules \cite{aot,mot}, see also \cite{km,kmwa} and the references cited therein. The immobilized molecules have not disappeared, they can further interact with the environment. In this case, the probability of finding a molecule in the system is equal to one at any time. We call the process subdiffusion with particle immobilization. It is obvious that this process cannot be described by a diffusion--reaction equation. 

The immobilization of molecules can occur in a medium in which the movement of particles is very hindered, as in the biofilm mentioned above, subdiffusion may occur in such a system, see for example Refs. \cite{mk,mk1,mks,bg,ks,klages2008,skb,sk,barkai2000,barkai2012}. We derive an equation describing subdiffusion with particle immobilization in a one-dimensional homogeneous system. We assume that after each jump a particle can be immobilized with the same probability which does not change with time and is independent of the particle position.

\section{Model}

To derive the subdiffusion--immobilization equation we use the continuous time random walk (CTRW) model \cite{mk,mk1,mks,skb,barkai2000,ks,montroll1965,compte,hilferanton,chechkin}. Within the model, when the average length of a single particle jump $\epsilon$ is finite the form of the subdiffusion equation is determined by the probability density $\psi$ of the waiting time for the particle to jump. 
In terms of the Laplace transform, $\mathcal{L}[f(t)](s)=\int_0^\infty {\rm e}^{-st}f(t)dt\equiv \hat{f}(s)$, the equation generated by the function $\psi$ is as follows
\begin{equation}\label{eq4}
s\hat{P}(x,s)-P(x,0)=\frac{\epsilon^2 s\hat{\psi}(s)}{2[1-\hat{\psi}(s)]}\frac{\partial^2 \hat{P}(x,s)}{\partial x^2},
\end{equation}
the derivation of this equation is described in Appendix.

We make the following assumptions:
\begin{enumerate}
	\item the probability of finding a particle in the system is equal to one at any time,
	\begin{equation}\label{eq2}
	\int_{-\infty}^\infty P(x,t)dx=1,
	\end{equation}
	\item since the particle can be permanently stopped, the probability that the particle will make a jump is less than one,
	\begin{equation}\label{eq3}
	\int_0^\infty \psi(t)dt<1.
	\end{equation}
\end{enumerate}

\subsection{Subdiffusion equation}

To obtain the subdiffusion equation we assume
\begin{equation}\label{eq5}
\hat{\psi}(s)=\frac{1}{1+\tau s^\alpha},
\end{equation}
$0<\alpha<1$, where $\tau$ is a parameter with the units of $\text{s}^\alpha$. This function satisfies the normalization condition 
\begin{equation}\label{eq6}
\int_0^\infty \psi(t)dt\equiv \hat{\psi}(0)=1.
\end{equation}
This condition means that the particle cannot be stopped permanently with non-zero probability.
From Eqs. (\ref{eq4}) and (\ref{eq5}) we get
\begin{equation}\label{eq7}
s\hat{P}(x,s)-P(x,0)=Ds^{1-\alpha}\frac{\partial^2 \hat{P}(x,s)}{\partial x^2},
\end{equation}
where $D=\epsilon^2/2\tau$ is a subdiffusion coefficient given in the units of $\text{m}^2/\text{s}^\alpha$. 
Due to the relations
\begin{equation}\label{eq8}
\mathcal{L}^{-1}\left[s\hat{f}(s)-f(0)\right](t)=\frac{df(t)}{dt},
\end{equation}
\begin{equation}\label{eq9}
\mathcal{L}^{-1}\left[s^{\beta} \hat{f}(s)\right](t)=\frac{^{RL}d^{\beta} f(t)}{dt^{\beta}},
\end{equation}
$0<\beta<1$, where 
\begin{equation}\label{eq10}
\frac{^{RL}d^{\beta} f(t)}{d t^{\beta}}=\frac{1}{\Gamma(1-\beta)}\frac{d}{dt}\int_0^t (t-u)^{-\beta}f(u)du
\end{equation}
is the Riemann--Liouville time fractional derivative of the order $\beta\in(0,1)$.
From Eqs. (\ref{eq7})--(\ref{eq9}) we get the subdiffusion equation
\begin{equation}\label{eq11}
\frac{\partial P(x,t)}{\partial t}=D\frac{^{RL}\partial^{1-\alpha}}{\partial t^{1-\alpha}}\frac{\partial^2 P(x,t)}{\partial x^2}.
\end{equation}

\subsection{Subdiffusion--immobilization equation}

In order to find a function $\psi(t)$ that satisfies Eq. (\ref{eq3}), i.e. $\hat{\psi}(0)<1$, we assume that the Laplace transform of the function is 
\begin{equation}\label{eq12}
\hat{\psi}(s)=\frac{1}{1+\tau\gamma+\tau s^\alpha},
\end{equation}
$0<\alpha<1$, the parameter $\gamma$, which controls molecule immobilization, is given in the units of $1/\text{s}^\alpha$. The probability $p_s$ of stopping the molecule permanently is $p_s=1-\hat{\psi}(0)=\tau\gamma/(1+\tau\gamma)$. From Eqs. (\ref{eq4}) and (\ref{eq12}) we get
\begin{equation}\label{eq13}
s\hat{P}(x,s)-P(x,0)=D\frac{s^{1-\alpha}}{1+\gamma s^{-\alpha}}\frac{\partial^2 \hat{P}(x,s)}{\partial x^2}
\end{equation}
The inverse Laplace transform of the right-hand side of Eq. (\ref{eq13}) is calculated using the formula
\begin{equation}\label{eq14}
\mathcal{L}^{-1}\left[\frac{s^{1-\alpha}}{1+\gamma s^{-\alpha}}\hat{f}(s)\right](t)=\frac{^{RL}_{\;\;F} d^{1-\alpha} f(t)}{dt^{1-\alpha}},
\end{equation}
where 
\begin{equation}\label{eq15}
\frac{^{RL}_{\;\;F} d^{1-\alpha} f(t)}{d t^{1-\alpha}}=\frac{d}{dt}\int_0^t F_\alpha (t-t';\gamma)f(t')dt'
\end{equation}
is the Riemann--Liouville type fractional derivative with the kernel $F_\alpha$ which is defined by its Laplace transform
\begin{equation}\label{eq16}
\hat{F}_\alpha (s;\gamma)=\frac{1}{\gamma+s^\alpha}.
\end{equation}
For $\gamma=0$, this derivative is the Riemann--Liouville derivative Eq. (\ref{eq10}) of the order $1-\alpha$. 
Eqs. (\ref{eq13})--(\ref{eq16}) provide the following subdiffusion--immobilization equation
\begin{equation}\label{eq17}
\frac{\partial P(x,t)}{\partial t}=D\frac{^{RL}_{\;\;F}\partial^{1-\alpha}}{\partial t^{1-\alpha}}\frac{\partial^2 P(x,t)}{\partial x^2}.
\end{equation}

Calculation of the inverse transform of Eq. (\ref{eq16}) is usually done by power series expansion of the function when $\gamma/s^\alpha<1$, and then inverting the transform term by term using the formula $\mathcal{L}^{-1}[1/s^\beta](t)=t^{\beta-1}/\Gamma(\beta)$, $\beta>0$. The result is the Mittag-Leffler function \cite{mainardi1,mainardi2}. However, this procedure is valid for relatively large values of the parameter $s$, which correspond to small values of time variable. To get the inverse Laplace transform over the whole time domain we propose to use the following method: (1) instead of $\hat{F}_\alpha$ Eq. (\ref{eq16}) find the inverse transform of $\hat{F}_\alpha(s,\gamma){\rm e}^{-as^\mu}$, $a,\mu>0$, (2) expand $\hat{F}_\alpha$ in a power series of $s$ considering both cases $s^\alpha>\gamma$ and $s^\alpha<\gamma$ separately, (3) use the formula \cite{tkoszt2004}
\begin{eqnarray}\label{eq18}
\mathcal{L}^{-1}\left[s^\nu{\rm e}^{-as^\mu}\right](t)\equiv f_{\nu,\mu}(t;a)\\
=\frac{1}{t^{\nu+1}}\sum_{n=0}^\infty \frac{1}{n!\Gamma(-n\mu-\nu)}\left(-\frac{a}{t^\mu}\right)^n\nonumber
\end{eqnarray}
$a,\mu>0$, (4) calculate the limit of $a\rightarrow 0^+$ in the obtained functions. We note that
\begin{equation}\label{eq19}
f_{\nu,\mu}(t;0^+)=\frac{1}{t^{\nu+1}\Gamma(-\nu)},
\end{equation}
and the result is independent of the parameter $\mu$.

\begin{figure}[htb]
\centering{%
\includegraphics[scale=0.4]{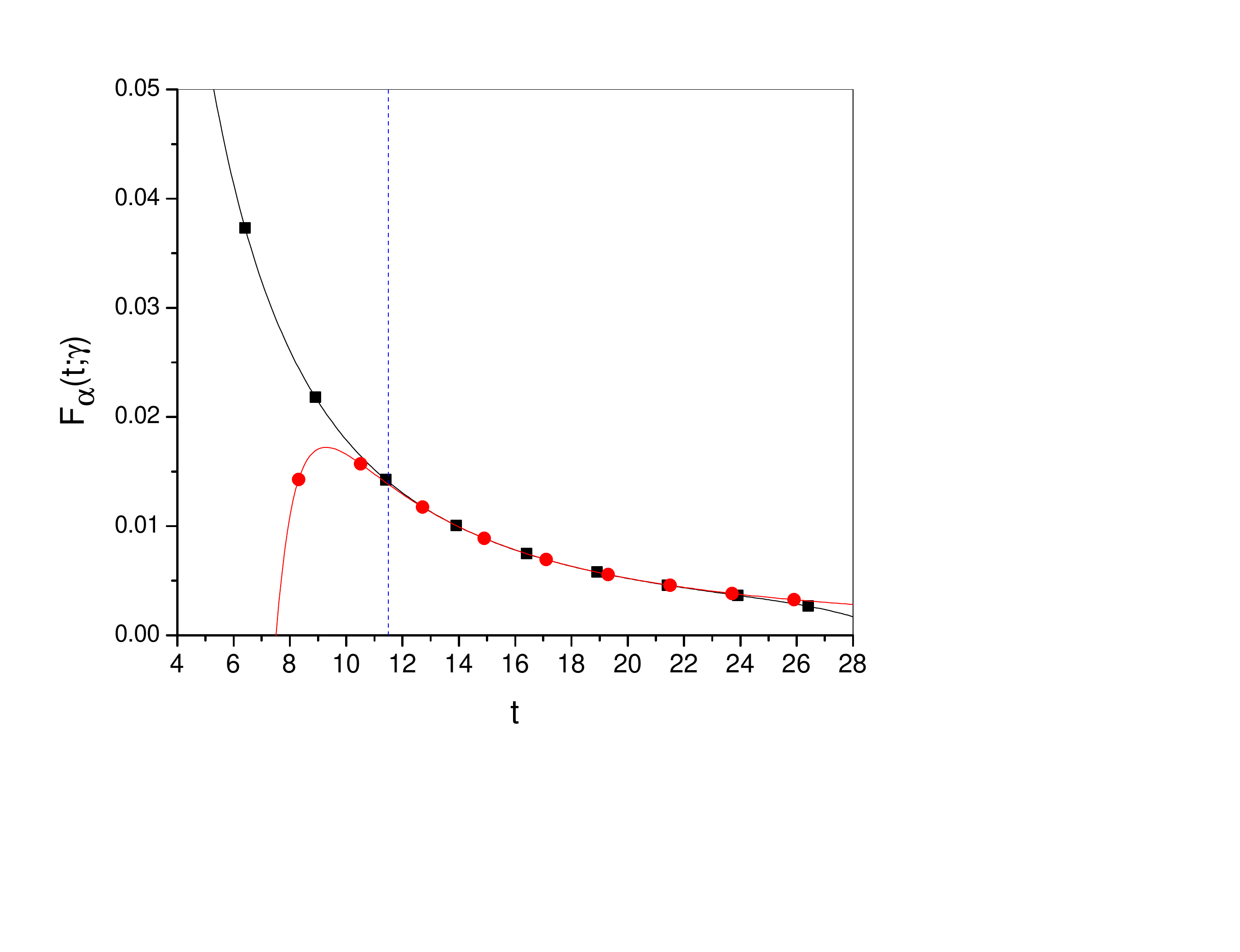}}
\caption{Plot of the function $F_\alpha$. The dashed vertical line shows the location of the parameter $t_b=11.5$. The solid line with squares is the plot of the upper function in Eq. (\ref{eq21}) which describes $F_\alpha$ for $t<t_b$, the solid line with circles is the plot of the lower function in Eq. (\ref{eq21}) which represents $F_\alpha$ for $t>t_b$. In the numerical calculations, the leading 20 terms in the series appearing in the functions $E_{\alpha,\alpha}$ and $\tilde{E}_{\alpha,\alpha}$ have been taken into account.}
\label{fig1}
\end{figure}

From the formula
\begin{eqnarray}\label{eq20}
\frac{{\rm e}^{-as^\mu}}{\gamma+s^\alpha}=
\left\{
\begin{array}{c}
{\rm e}^{-as^\mu}\sum_{n=0}^\infty (-\gamma)^n s^{-(n+1)\alpha},\;s>\gamma^{1/\alpha},\\
   \\
\frac{{\rm e}^{-as^\mu}}{\gamma}\sum_{n=0}^\infty \left(-\frac{1}{\gamma}\right)^n s^{n\alpha},\;s<\gamma^{1/\alpha},
\end{array}
\right.
\end{eqnarray}
and Eqs. (\ref{eq18}) and (\ref{eq19}) we obtain
\begin{eqnarray}\label{eq21}
F_\alpha(t;\gamma)=
\left\{
\begin{array}{c}
\frac{1}{t^{1-\alpha}}E_{\alpha,\alpha}(-\gamma t^\alpha),\;t<t_b,\\
   \\
-\frac{1}{\gamma^2 t^{1+\alpha}}\tilde{E}_{\alpha,\alpha}\left(-\frac{1}{\gamma t^\alpha}\right),\;t>t_b,
\end{array}
\right.
\end{eqnarray}
where $E_{\alpha,\beta}(u)=\sum_{n=0}^\infty \frac{u^n}{\Gamma(\alpha n+\beta)}$, $\alpha,\beta>0$, is the two--parameter Mittag--Leffler (ML) function, $\tilde{E}_{\alpha,\beta}(u)=\sum_{n=0}^\infty \frac{u^n}{\Gamma(-\alpha n-\beta)}$ is a generalization of the ML function for negative parameters. We note that conditions $s>\gamma^{1/\alpha}$ and $s<\gamma^{1/\alpha}$ do not determine the parameter $t_b$. For example, the condition $s>\gamma^{1/\alpha}$ is equivalent to $1/s^{\beta+1}<1/(s^\beta\gamma^{1/\alpha})$ for $\beta>0$ (assuming that $s$ is a real positive parameter). The inverse Laplace transform of the inequality provides $t<\beta/\gamma^{1/\alpha}$ where $\beta$ is a positive number. Thus, the above inequality does not determine $t_b$. Here we define the parameter $t_b$ as the shorter time at which the upper and the lower functions in Eq. (\ref{eq21}) are matched, see Fig. \ref{fig1}.

In terms of the Laplace transform the solution to Eq. (\ref{eq16}) (the Green's function) for the initial condition $P(x,0)=\delta(x)$, where $\delta$ is the Dirac--delta function, and boundary conditions $P(\pm\infty,t)=0$ is
\begin{equation}\label{eq22}
\hat{P}(x,s)=\frac{\sqrt{\gamma+s^\alpha}}{2s\sqrt{D}}\;{\rm e}^{-|x|\frac{\sqrt{\gamma+s^\alpha}}{\sqrt{D}}}
\end{equation}
The solution fulfils the condition $\int_{-\infty}^\infty\hat{P}(x,s)dx=1/s$ what provides the normalization of the function $P$ Eq. (\ref{eq2}). 

Let $\gamma\neq 0$. We calculate the inverse Laplace transform of the function (\ref{eq22}) for small and large values of $s$ separately. 
In calculation, we use the formulas $\sqrt{1+u}\approx 1+u/2-u^2/8$ and ${\rm e}^{-u}\approx 1-u+u^2/2$, $u\rightarrow 0$, and keep the leading terms in the obtained series. When $s^\alpha>\gamma$ we obtain
\begin{equation}\label{eq23}
\hat{P}(x,s)=\frac{1}{2\sqrt{D}s^{1-\alpha/2}}\left(1-\frac{b_1}{s^{\alpha/2}}+\frac{b_2}{s^\alpha}\right){\rm e}^{-\frac{|x|}{\sqrt{D}}s^{\alpha/2}},
\end{equation}
where $b_1=\gamma|x|/2\sqrt{D}$ and $b_2=(\gamma/2)(1+|x|^2\gamma/2\sqrt{D})$.
If $s^\alpha<\gamma$, we get
\begin{equation}\label{eq24}
\hat{P}(x,s)=\frac{\sqrt{\gamma}}{2s\sqrt{D}}{\rm e}^{-\sqrt{\frac{\gamma}{D}}|x|(1+\frac{s^\alpha}{2\gamma})}\left[1+\frac{s^\alpha}{2\gamma}-b\frac{s^{2\alpha}}{\gamma^2}\right],
\end{equation}
where $b=\sqrt{\gamma/D}|x|+1/8$.
Eqs. (\ref{eq18}) and (\ref{eq23}) provide the Green's functions in the limit of short time
\begin{eqnarray}\label{eq25}
P(x,t)=\frac{1}{2\sqrt{D}}\Big[f_{-1+\alpha/2,\alpha/2}(t;\eta)\\
-b_1 f_{-1,\alpha/2}(t;\eta)+b_2 f_{-1-\alpha/2,\alpha,2}(t;\eta)\Big],\nonumber
\end{eqnarray}
where $\eta=|x|/\sqrt{D}$. From Eqs. (\ref{eq18}) and (\ref{eq24}) we get the Green's function in the long time limit
\begin{eqnarray}\label{eq26}
P(x,t)=\frac{1}{2}\sqrt{\frac{\gamma}{D}}{\rm e}^{-\sqrt{\frac{\gamma}{D}}|x|}\Big[f_{-1,\alpha}(t;\xi)\\
+\frac{1}{2\gamma}f_{\alpha-1,\alpha}(t;\xi)-\frac{b}{\gamma^2}f_{2\alpha-1,\alpha}(t;\xi)\Big],\nonumber
\end{eqnarray}
where $\xi=|x|/2\sqrt{D\gamma}$.

Since the mean particle position equals zero, in terms of the Laplace transform the mean square displacement of the particle is
\begin{equation}\label{eq27}
\mathcal{L}\left[\left\langle \right(\Delta x)^2(t)\rangle\right](s)=\int_{-\infty}^\infty x^2\hat{P}(x,s)dx=\frac{2D}{s(\gamma+s^\alpha)}.
\end{equation}
When $\gamma\neq 0$, for small $s$ we have $\mathcal{L}\left[\left\langle \right(\Delta x)^2(t)\rangle\right](s)=2D/[1/s-1/(\gamma s^{1-\alpha})]$. Thus, in the limit of long time we get
\begin{equation}\label{eq28}
\left\langle (\Delta x)^2(t)\right\rangle=\frac{2D}{\gamma}\left[1-\frac{1}{\gamma\Gamma(1-\alpha)t^\alpha}\right].
\end{equation}

In the limit $t\rightarrow\infty$, the stationary state described by the following function is reached,
\begin{equation}\label{eq29}
P(x,t\rightarrow\infty)\equiv P_{st}(x)=\frac{1}{2}\sqrt{\frac{\gamma}{D}}\;{\rm e}^{-\sqrt{\frac{\gamma}{D}}|x|}.
\end{equation}

\begin{figure}[htb]
\centering{%
\includegraphics[scale=0.4]{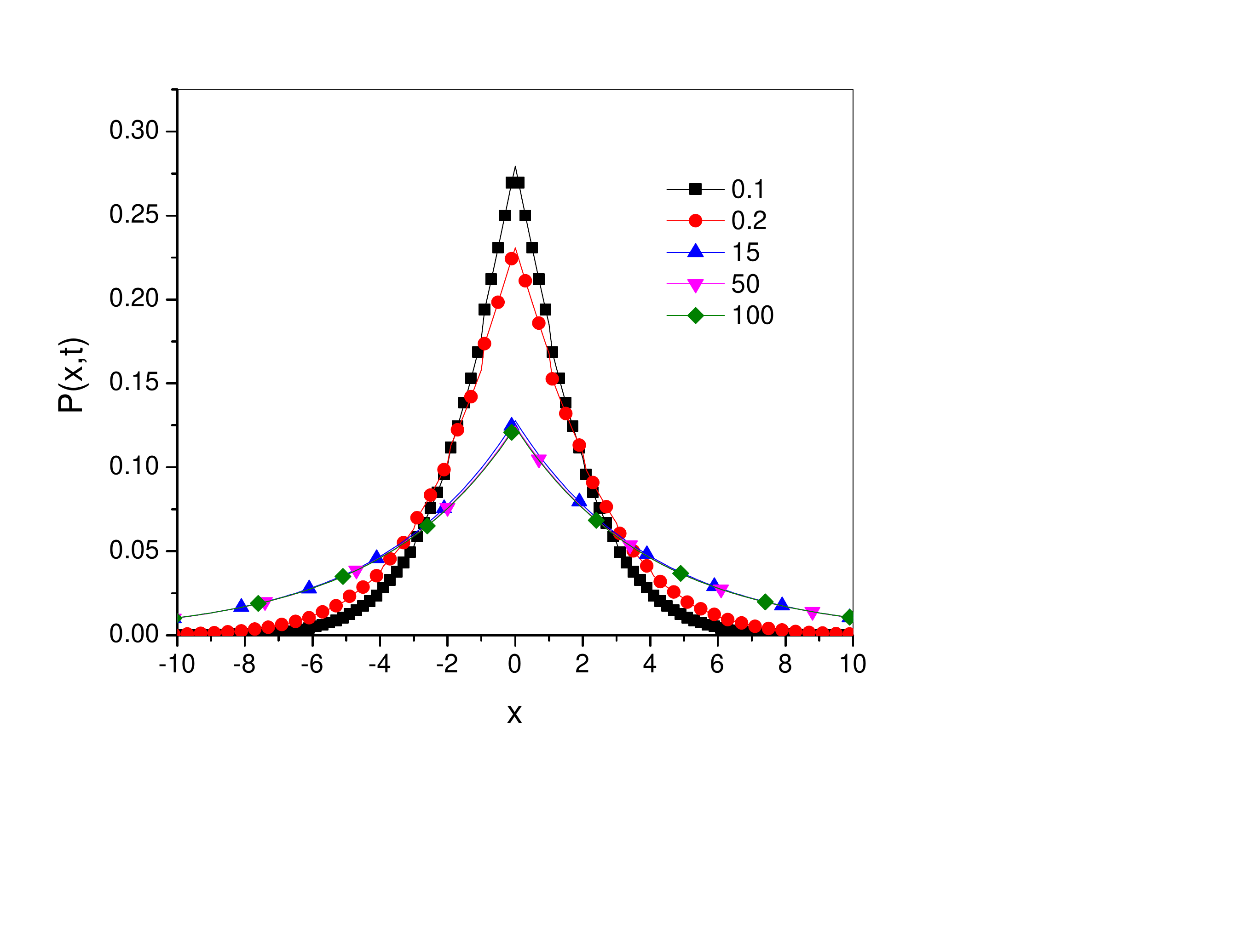}}
\caption{Plots of Green's functions for times given in the legend. The plots represent the function Eq. (\ref{eq23}) for $t=0.1, 0.5$ and Eq. (\ref{eq24}) for $t=15,50,100$.}
\label{fig2}
\end{figure}

\begin{figure}[htb]
\centering{%
\includegraphics[scale=0.4]{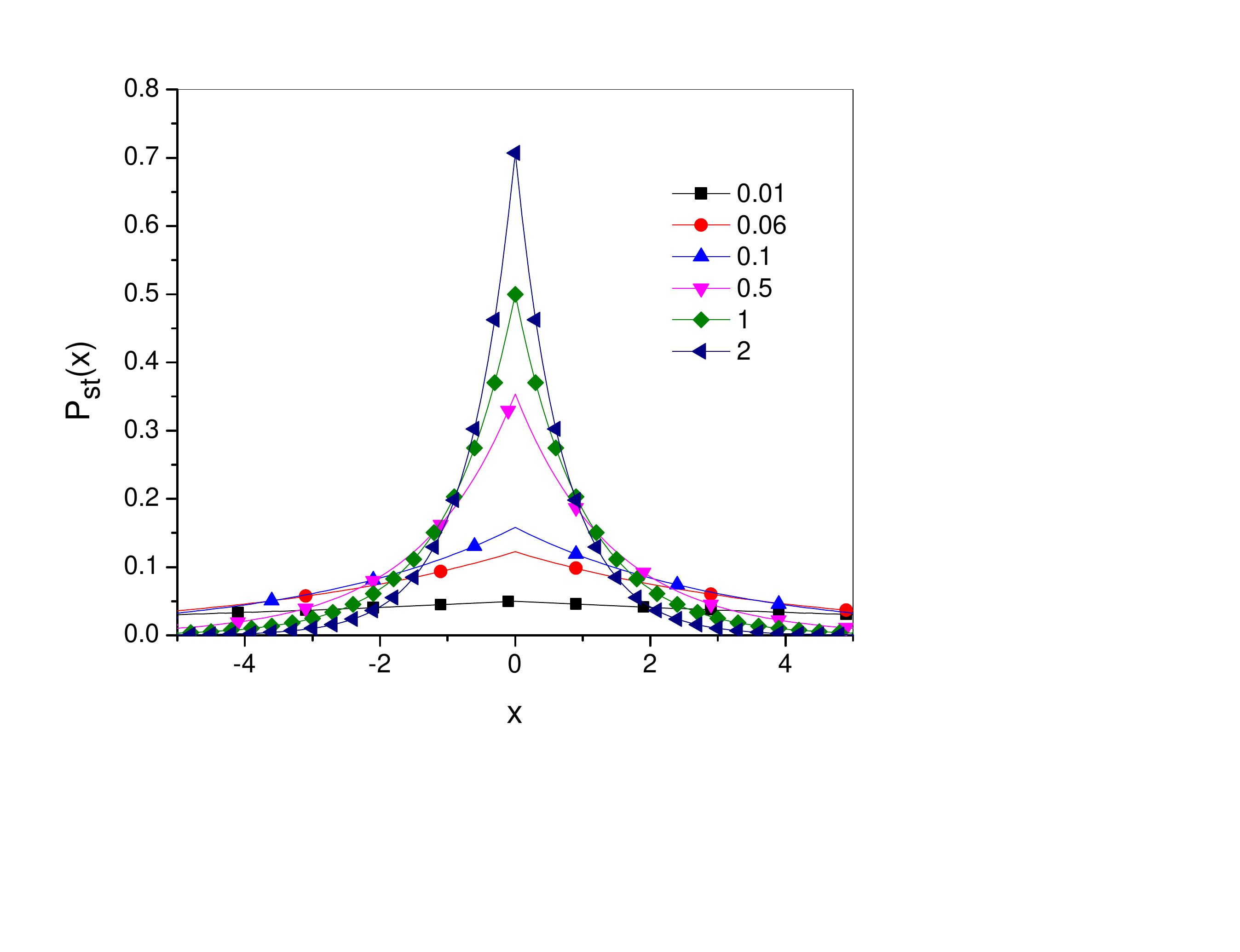}}
\caption{Plots of the function $P_{st}$ Eq. (\ref{eq29}) for different values of the ratio $\gamma/D$ given in the legend.}
\label{fig3}
\end{figure}

For illustration, plots of functions $F_\alpha$ and $P$ are shown in Figs. \ref{fig1} and \ref{fig2}, respectively. The parameters are $\alpha=0.7$, $\gamma=0.6$, and $D=10$, all parameters are given in arbitrarily chosen units. In Fig. \ref{fig3} the Green's functions for the stationary state are presented.

\section{Final remarks}

The process of subdiffusion with particle immobilization can be described by an equation with a fractional time derivative of the Riemann--Liouville type, which is a differential-integral operator with the kernel $F_\alpha$ defined by its Laplace transform Eq. (\ref{eq16}). Normal diffusion and subdiffusion have a different stochastic interpretation. However, the normal diffusion-immobilization equation can be obtained from Eq. (\ref{eq17}) by substituting $\alpha=1$. We have proposed a method for determining the inverse Laplace transform of the kernel. In our opinion, this method can be widely used for calculating inverse Laplace transforms $\mathcal{L}^{-1}[\hat{f}(s)](t)$ for a wide class of functions $f$. 

In a homogeneous unbounded system the subdiffusion-immobilization process reaches a stationary state which is described by $P_{st}(x)$ Eq. (\ref{eq29}). This distribution depends only on the quotient $\gamma/D$ expressed in the units of $1/\text{m}^2$ and it does not explicitly depend on the parameter $\alpha$. The achievement of the steady state is suggested by Fig. \ref{fig2}, where the Green's functions for relatively long times differ very little from each other. In the stationary state there is $\left\langle (\Delta x)^2(t\rightarrow\infty)\right\rangle=\frac{2D}{\gamma}$, the particle is finally immobilized with probability equal to one.

The subdiffusion--immobilization process is described by Eq. (\ref{eq17}) that can be obtained in practice by replacing the time fractional Riemann-Liouville derivative Eq. (\ref{eq10}) with the more general Riemann--Liouville type derivative with the kernel $F_\alpha$ Eq. (\ref{eq15}) in the ``ordinary'' subdiffusion equation Eq. (\ref{eq11}), orders of both derivatives are the same. There is a different situation than in the subdiffusion--reaction equation in which the reaction term is involved in the ``ordinary'' subdiffusion equation, see for example Refs. \cite{mendez,yuste,kl2014}. We mention that the Riemann--Liouville type fractional derivatives with different kernels have been considered in Ref. \cite{hl1998,lh1999,lh2000,yang}.

\section*{Acknowledgment}

The author wishes to express his thanks to Aldona Dutkiewicz for fruitful discussions.

\section*{Appendix. Derivation of Eq. (\ref{eq4})}

To derive the subdiffusion equation we use a simple model of a particle random walk along a one--dimensional homogeneous lattice. 
Usually, in the CTRW model both a particle jump length and waiting time for a particle to jump are random variables. We assume that the jump length distribution $\lambda$ has the form $\lambda(x)=\frac{1}{2}[\delta(x-\epsilon)+\delta(x+\epsilon)]$. Random walk with discrete time $n$ is described by the equation $P_{n+1}(m)=\frac{1}{2}P_n(m+1)+\frac{1}{2}P_n(m-1)$, where $P_n(m)$ is a probability that a diffusing particle is at the position $m$ after $n$-th step. Let the initial particle position be $m=0$. Moving from discrete $m$ to continuous $x$ spatial variable we assume $x=m\epsilon$ and $P_n(x)=P_n(m)/\epsilon$, where $\epsilon$ is a distance between discrete sites. The above equations and the relation $[P_n(x+\epsilon)+P_n(x-\epsilon)-2P_n(x)]/\epsilon^2 =\partial^2 P_n(x)/\partial x^2$, $\epsilon\rightarrow 0$, provide the following equation in the limit of small $\epsilon$
\begin{equation}\label{a1}
P_{n+1}(x)-P_n(x)=\frac{\epsilon^2}{2}\frac{\partial^2 P_n(x)}{\partial x^2}.
\end{equation}
To move from discrete to continuous time we use the formula $P(x,t)=\sum_{n=0}^\infty Q_n(t)P_n(x)$ \cite{montroll1965}, 
where $Q_n(t)$ is the probability that a diffusing particle takes $n$ step in the time interval $(0,t)$. The function $Q_n$ is a convolution of $n$ distributions $\psi$ of a waiting time for a particle to jump and a function $U(t)=1-\int_0^t \psi(t')dt'$ which is the probability that a particle does not change its position after $n$-th step, $\hat{U}(s)=[1-\hat{\psi}(s)]/s$, $Q_n(t)=(\underbrace{\psi\ast\psi\ast\ldots\ast\psi}_{n\;times}\ast U)(t)$, where $(f\ast h)(t)=\int_0^t f(u)h(t-u)du$. Due to the following property $\mathcal{L}[(f\ast h)(t)](s)=\hat{f}(s)\hat{h}(s)$ we obtain
\begin{equation}\label{a2}
\hat{P}(x,s)=\frac{1-\hat{\psi}(s)}{s}\sum_{n=0}^\infty \hat{\psi}^n (s) P_n(x).
\end{equation}
Combining Eqs. (\ref{a1}) and (\ref{a2}) we get Eq. (\ref{eq4}).


\begin{thebibliography}{33}

\bibitem{aot} G. G. Anderson and G. A. O'Toole, Bacterial Biofilms, Current Topics in Microbiology and Immunology {\bf 322} (Berlin, Springer, 2008).
\bibitem{mot} T. F. C. Mah and G. A. O'Toole, Mechanisms of biofilm resistance to antimicrobial agents, Trends Microbiol. {\bf 9}, 34 (2001).
\bibitem{km} T. Koszto{\l}owicz and R. Metzler, Diffusion of antibiotics through a biofilm in the presence of diffusion and absorption barriers, Phys. Rev. E {\bf 102}, 032408 (2020).
\bibitem{kmwa} T. Koszto{\l}owicz, R. Metzler, S. W\c{a}sik, and M. Arabski, Modelling experimentally measured of ciprofloxacin antibiotic diffusion in  Pseudomonas aeruginosa biofilm formed in artificial sputum medium, PLoS ONE {\bf 15}(12), e0243003 (2020).
\bibitem{mk} R. Metzler and J. Klafter, The random walk's guide to anomalous diffusion: a fractional dynamics approach, Phys. Rep.  {\bf 339}, 1 (2000).
\bibitem{mk1} R. Metzler and J. Klafter, The restaurant at the end of the random walk: recent developments in the description of anomalous transport by fractional dynamics, J. Phys. A {\bf 37}, R161 (2004).
\bibitem{mks} R. Metzler, J. Klafter, and I. M. Sokolov, Anomalous transport in external fields: Continuous time random walks and fractional diffusion equations extended, Phys. Rev. E {\bf 58}, 1621 (1998).
\bibitem{bg} J. P. Bouchaud and A. Georgies, Anomalous diffusion in disordered media: statistical mechanisms, models and physical applications, Phys. Rep. {\bf 195}, 127 (1990).
\bibitem{ks} J. Klafter and I. M. Sokolov, {\it First Step in Random Walks. From Tools to Applications} (Oxford UP, New York, 2011).
\bibitem{klages2008} R. Klages, G. Radons, and I. M. Sokolov, {\it Anomalous Transport: Foundations and Applications} (Wiley, New York, 2008).
\bibitem{skb} I. M. Sokolov, J. Klafter, and A. Blumen, Fractional kinetics, Phys. Today {\bf 55}, 11, 48-54 (2002).
\bibitem{sk} I. M. Sokolov and J. Klafter, From diffusion to anomalous diffusion: a century after Einstein's Brownian motion, Chaos {\bf 15}, 026103 (2005).
\bibitem{barkai2000} E. Barkai, R. Metzler, and J. Klafter, From continuous time random walks to the fractional Fokker-Planck equation, Phys. Rev. E {\bf 61}, 132 (2000).
\bibitem{barkai2012} E. Barkai, Y. Garini, and R. Metlzer, Strange kinetics of single molecules in living cells, Phys. Today \textbf{65}, 29 (2012).
\bibitem{montroll1965} E. W. Montroll and G. H. Weiss, Random walks on lattices. II, J. Math. Phys. {\bf 6}, 167 (1965).
\bibitem{compte} A. Compte, Stochastic foundations of fractional dynamics, Phys. Rev. E {\bf 53}, 4191 (1996).
\bibitem{hilferanton} R. Hilfer and L. Anton, Fractional master equations and fractal time random walks, Phys. Rev. E {\bf 51}, R848 (1995).
\bibitem{chechkin} A. V. Chechkin, M. Hofmann, and I. M. Sokolov, Continuous-time random walk with correlated waiting times, Phys. Rev. E {\bf 80}, 031112 (2009).
\bibitem{mainardi1} F. Mainardi, A tutorial on the basic special functions of fractional calculus, WSEAS Trans. Math. {\bf 19}, 74 (2020).
\bibitem{mainardi2} F. Mainardi, Why the Mittag-Leffler function can be considered the Queen function of the fractional calculus?, Entropy {\bf 22}, 1359 (2020).
\bibitem{tkoszt2004} T. Koszto{\l}owicz, From the solutions of diffusion equation to the solutions of subdiffusive one, J. Phys. A: Math. Gen. {\bf 37}, 10779 (2004).
\bibitem{mendez} V. M\'endez, S. Fedotov, and W. Horsthemke, {\it Reaction--Transport Systems: Mesoscopic Foundations, Fronts, and Spatial Instabilities} (Springer, Berlin, 2010).
\bibitem{yuste} S.B. Yuste, L. Acedo, and K. Lindenberg, Reaction front in an $A+B\rightarrow C$ reaction-subdiffusion process, Phys. Rev. E {\bf 69}, 036126 (2004).
\bibitem{kl2014} T. Koszto{\l}owicz and K.D. Lewandowska, Subdiffusion-reaction processes with $A\rightarrow B$ reactions versus subdiffusion-reaction processes with $A+B\rightarrow B$ reactions, Phys. Rev. E \textbf{90}, 032136 (2014).
\bibitem{hl1998} T.T. Hartley and C.F. Lorenzo, A solution to the fundamental linear fractional order differential equation, NASA/TP-1998-208693 (1998).
\bibitem{lh1999} C.F. Lorenzo and T.T. Hartley, Generalized functions for the fractional calculus, NASA/TP-1999-209424/REV1 (1999).
\bibitem{lh2000} C.F. Lorenzo and T.T. Hartley, R--function relationships for application in the fractional calculus, NASA/TM-2000-210361 (2000).
\bibitem{yang} X.-J. Yang, {\it General Fractional Derivatives. Theory. Methods and Applications}, CRC Press, Taylor and Francis Group, Boca Raton (2019), p.177.

\end{thebibliography}
\end{document}